# Current induced forces upon atoms adsorbed on conducting carbon nanotubes

N.Mingo*[(1)], Liu Yang[(2)] and Jie Han[(2)]

[1]NASA-Ames res.Center,
T27A-1 Moffett Field, CA94035

[2]Eloret Corp.
NASA-Ames res.Center,
Moffett Field, CA94035

* E-mail: mingo@nas.nasa.gov

**Abstract**

We calculate the forces acting upon species adsorbed on a single wall carbon nanotube, in the presence of electric currents. We present a self consistent real space Green function method, which enables us to calculate the current induced forces from an ab-initio Hamiltonian. The method is applied to calculate the force on an adsorbed O atom on a (5,5) carbon nanotube, for different bias voltages and adsorption sites. For good contact regimes and biases of the order of Volts, the presence of a current can affect the potential energy surfaces considerably. Implications of these effects for the induced diffusion of the species are analyzed. The dependence of the force with the nanotube radius is studied. In addition, the magnitude of inelastic electron scattering, inducing vibrational heating, and its influence on the adsorbates' drift, is commented.

# I-Introduction

A great ammount of work is currently being done in the field of molecular electronics, amongst other reasons because of its promising technological applications. Computational modeling plays an important role in understanding the basic mechanisms involved in the functioning of these atomic scale devices. From both physical[1] and chemical[2] points of view, computational techniques have emerged, capable of predicting stable atomic structures of systems in their ground state. Quantitative calculation of tunneling currents at nanoscale contacts is also evolving towards enhanced accuracy, by combining methods from quantum chemistry and solid state physics[3,4]. One important issue concerns the combination of the previous two, namely the effect of the tunneling currents on the stable structures of these atom sized systems. Despite an important ammount of work has been done on the topic of impurities' electromigration in extended systems[5,6], very few studies exist focusing on the forces induced by current at molecular wires or contacts. A recent paper by Todorov et al. deals with the current induced forces on metalic s-orbital nanowires[7]. Here we treat a problem of direct importance for the molecular electronics field: the current induced forces on the adsorbed species at a carbon nanotube conducting wire. In addition to the basic science as well as methodological interest, this phenomenon has technological importance due to its relation with nanodevice degradation[8,9]. The main questions we aim to answer here are: how big are the forces? What is their direction as a function of the current? Which effects can be predicted in relation to induced diffusion? All these might be of importance, for instance, in recently demonstrated controlled doping of nanotube based p-n junctions[10a], where intense current flow during a long time might deteriorate the dopants' spatial distribution. Another interesting application might be the controlled drift of adsorbates along the nanotube's surface, as recently proposed[10b].

The drift induced on adsorbates by the flowing electrons is caused by two different processes, namely elastic and inelastic. The elastic processes have largely been the subject of electromigration theories. Elastic scattering of the electrons by an impurity leads to the self-consistent redistribution of charges in the system when there is a stationary current. This modified electron distribution induces a force on the atoms of the system, due to electrons transfering momentum to the atoms. One can consider that due to these forces the atomic positions adiabatically shift towards new equilibrium positions. This in turn modifies the potential energy surface of the impurity, and its diffusion barriers, leading to an enhanced mobility in a particular direction. In electromigration, these forces are usually divided in external and wind contributions, however their origin is the same, and such a distinction is not essential for calculations[7]. The derivation of the expression for the steady forces, which are the mean value of the electronic force due to the flowing electrons redistribution, can be obtained from a Hellman-Feynman type expression[12c].

There is a difference between the steady state forces, and those induced by inelastic scattering of electrons. To see this let us consider the following example. Suppose an adatom adsorbed on a bridge position between two symmetrical substrate atoms, and now let a current flow along the surface, in the direction perpendicular to the bridge. By symmetry, the static force cannot have any component parallel to the bridge. However, excitation of the adsorbate's vibrational mode parallel to the bridge is still possible due to fluctuations of the charge density (a model calculation of this has been previously shown[11]). In other words, despite the mean value of the force component is 0, the mean value of its modulus is different than 0.

The inmediate consequence of this, is that inelastic scattering of flowing electrons, transfering energy to the adsorbate's vibrational modes, can excite the adsorbates' vibrational states to a level with enough energy to overcome the diffusion barrier. The population of the adsorbates' vibrational levels as a function of the current in the incoherent regime has been formulated by Persson and Avouris[9] in terms of an effective adsorbate temperature, different than the sample's temperature. Such formulation allows to calculate the enhanced hopping rates. In this paper we focus on the calculation of the static force induced by currents, briefly commenting on the inelastic scattering processes in section V.

The structure of the paper is as follows. In section II we explain the theoretical techniques we use to calculate the forces when a current is present, as well as the way to compute the non-equilibrium electronic distribution. Then (sec.III) we summarize the features of the computational techniques employed, and the system's configuration. The results' section (IV) afterwards shows adsorption energy curves for atomic O, then the electronic charge redistribution is studied as a function of the applied bias, and the forces are obtained. Possibilities of induced directional diffusion are discussed, and a comment is made on inelastic scattering processes. The force dependence with radius is studied afterwards. A summary of results and conclusions is given in sec.V.

## II.-Theory of current induced mean forces

The Hellman-Feynman theorem[12] states that the change in the total energy of a quantum mechanical system upon variation of a parameter $\alpha$ equals the expected value of the derivative of the Hamiltonian with respect to $\alpha$. When $\alpha$ is an atomic coordinate, this means that the component of the force on the atom in that direction equals the classical force exerted by the electronic cloud corresponding to the expected value of the charge density at every point.

If we calculate the charge redistribution when a voltage is applied and a current is allowed to flow through the tube, the force induced on the coordinate $\alpha$ can be obtained (see note[7]). Considering a local atomic basis this takes the form:

$$(1) \quad F_\alpha \equiv \frac{d}{d\alpha}E = \sum_{ij} \delta q_{ij} \frac{dH_{ij}}{d\alpha} \quad ,$$

where $\delta \mathbf{q}$ is the change in the density matrix of the system after applying the current. A careful derivation of this expression is given by Todorov et al[7]. Therefore, we just need to compute the non-equilibrium charge density in the presence of current flow, and from it obtain the force on the atom using eq.(1), where $\alpha$ is the coordinate along which the force is calculated. The charge density is obtained from the non-equilibrium Green function in the following manner:

$$(2) \quad q_{ij} = \int_{-\infty}^{\infty} f(E)\rho_{ij}^{tot}(E)dE + \int_{-\infty}^{\infty} [f(E-eV) - f(E)]\rho_{ij}^{trans}(E)dE$$

Here the transmitted density $\rho^{trans}$ corresponds to the density of states that enter from the emitter. In terms of Green functions, the matrix for the transmitted l.d.o.s. at ring $i$ of the tube is

$$(3) \quad \underline{\rho}_i^{trans} = \mathbf{D}_i \underline{\rho}_i^0 \mathbf{D}_i^\dagger$$

where

$$(4) \quad \mathbf{D}_i = \left(\mathbf{I} - \mathbf{g}_i^l \mathbf{H}_{i,i+1} \mathbf{g}_{i+1}^r \mathbf{H}_{i+1,i}\right)^{-1} \quad , \quad \underline{\rho}_i^0 = \frac{1}{\pi} Im \mathbf{g}_i^l$$

$\mathbf{g}_i^l$ is the one electron retarded Green function matrix corresponding to layer $i$ of the system when layers $i$ and $i+1$ are uncoupled (l reminds that we are at the 'left' -or emitter- hand side of the division) and $\mathbf{g}_{i+1}^r$ is the green function at layer i+1 in the same divided situation (r reminds that we are at the 'right' hand side of the division). $\mathbf{H}_{i,i+1}$ is the Hamiltonian matrix coupling layer $i$ to $i+1$. The simplest way to derive eq.(3-4) is by using Lipman-Schwinger's equation to express the total-system (propagating) wave functions in terms of the wave functions of the uncoupled system at the sides of the interface between i and i+1. Let us denote the layers at the sides of the interface by 1 (emitter side) and 2 (collector side). Defining $\varphi$ as the wave functions for the decoupled system corresponding to $T_{12}=0$, and g the green functions for the uncoupled system, we have for the coupled system wave functions:

$$(5) \quad \Psi_n(1) = \varphi_n(1) + g_{11} T_{12} \Psi_n(2)$$

$$(6) \quad \Psi_n(2) = g_{22} T_{21} \Psi_n(1)$$

where we have used the fact that $g_{12}=0$, and $\varphi(2)=0$. These wave fucntions propagate from 1 to 2. In the same way, one can obtain the set of waves propagating from 2 to 1 by using those $\varphi_n$ such that $\varphi_n(1)=0$. Equations (5-6) imply

$$(7) \quad \Psi_n(1) = \left[ I - g_{11} T_{12} g_{22} T_{21} \right]^{-1} \varphi_n(1)$$

$$(8) \quad \Psi_n(2) = g_{22} T_{21} \left[ I - g_{11} T_{12} g_{22} T_{21} \right]^{-1} \varphi_n(1).$$

The l.d.o.s. corresponding to electrons propagating from 1 to 2 is thus

$$(9) \quad \underline{\rho}_1^{trans} = \sum_{n \in emitter} \left| \Psi_n(1) \right|^2 = \mathbf{D}_1 \underline{\rho}_1^{0} \mathbf{D}_1^{\dagger},$$

corresponding to eq.1. The result is expressed exclusively in terms of Green functions, so we do not need to consider the wave functions for calculational purposes. The total l.d.o.s. can be obtained directly from Dyson's equation:

$$(10) \quad G_{11} = g_{11} + g_{11} T_{12} G_{21} = g_{11} + g_{11} T_{12} g_{22} T_{21} G_{11} = \left[ I - g_{11} T_{12} g_{22} T_{21} \right]^{-1} g_{11}$$

$$(11) \quad \underline{\rho}_{11}^{total} = \frac{1}{\pi} Im(G_{11})$$

The above formulas enable to completely calculate the current-induced static forces on the adsorbate. (The calculation of the non-equilibrium charge density has to be done self-consistently, as we describe in ref.[13].) The procedure thus consists in defining one interface that splits the system in two parts, emitter side and collector side, which allows us to compute the total and transmitting l.d.o.s. at the emitter side. This allows to calculate the total charge in the system for a given applied bias, and do an iterative loop until self consistency is achieved. Then forces can be computed according to eq.(1). In the actual computation it is more efficient to successively define the interface at different places, which divide the system in different layers. In such a way we can avoid computing the green function connecting different layers. This is done via the following algorithm:

first one computes the self energy for the left and right electrodes, $\Sigma_l$, $\Sigma_r$ (see next paragraph). Let us consider the system divided in N layers. Defining $g_i^l$ as the Green function of the uncoupled system with the division interface placed between i and i+1, and $g_i^r$ as the one with the interface placed between i-1 and i, we calculate (see reference[14]):

$$(12) \quad g_1^l = [EI - H_{11} - \Sigma_l(E)]^{-1}$$

$$(13) \quad \Sigma_i^l(E) = H_{i, i-1} g_{i-1}^l H_{i-1, i}$$

$$(14) \quad g_i^l = [EI - H_{ii} - \Sigma_i^l(E)]^{-1}, i = 2, N$$

$$(15) \quad g_N^r = [EI - H_{NN} - \Sigma_r(E)]^{-1} = G_N$$

$$(16) \quad G_i = g_i^l + g_i^l H_{i, i+1} G_{i+1} H_{i+1, i} g_i^l, i = N-1, 1$$

$$(17) \quad \Sigma_i^r(E) = H_{i, i+1} g_{i+1}^r H_{i+1, i}$$

$$(18) \quad g_i^r = [EI - H_{ii} - \Sigma_i^r(E)]^{-1}, i = N-1, 1.$$

The trasnmitting and total l.d.o.s. are then calculated from eqs.(4,9) and (11).

There are different recursive methods to obtain the self-energy. Perhaps the most efficient one is the 'decimation technique'[15,16], which is based in the renormalization method, doubling the size of the system at each step. Convergence for each energy value is usually achieved in ~10 iterations or less, where only one matrix inversion is performed at each iteration, with matrices of the size of the repeated unit cell. The technique is described in the original reference[15], and also in Lanoo and Friedel[16]. However, implementation of the equations in the form provided by the latter led to convergence problems, which did not occur when using the original form of the equations by Guinea et al..

## III-Calculation details

First of all, we have calculated the O adsorption energy in the absence of current, by using the pseudopotential ab-initio method of Sankey et al.[17] ('fireball'). One essential feature of this method consists in the use of a cut-off radius for the wave functions, in such a way that they correspond to slightly excited states. The advantage of doing this is that the number of 3-center integrals to be computed is greatly reduced. The method has been tested for many different systems, showing remarkable accuracy[18]. In our case, a cut radius of 4.1 for C and 3.6 $a_0$ for O has been used, in accordance with previous calculations[19]. Adsorption energy curves were obtained as a function of the ad-atom's position, while the carbon atoms were assumed to be unrelaxed. A coverage of 1/8 O atoms per C was used. We have performed the ab-initio calculation of adsoprtion on an infinite graphene sheet, as a basis for the obtention of the L.C.A.O. Hamiltonian to be used for the force calculations. The same adsorption distances were assumed for all the tubes considered.

After obtaining the adsorption energy minimum, we have calculated the forces induced on the adsorbate when current is allowed to flow through the system, being the geometrical structure fixed at the adsorption configuration. To this end, the non-equilibrium Green functions and electron densities of the system have been computed self consistently in the mean field approximation. An accelerated convergence self-consistent algorithm is used for this purpose[20]. Green function calculation is computationally time consuming. Therefore, the computation has to be optimised by taking advantage of the tridiagonal character of the Hamiltonian (see sec.II), which avoids calculating Green function elements between disconnected layers. This procedure makes the computation to scale rather linearly with the system's size. Our Hamiltonian calculated by the ab-initio fireball approach of ref.[17] is in a non-orthogonal atomic orbital basis. In order to calculate the Green function, we further transform the Hamiltonian to an orthogonal basis, by a Lowdin transformation $H^{ort}=S^{-1/2}HS^{-1/2}$, where S is the overlap matrix. To speed up the Green function computation, interactions to second neighbors and beyond have been neglected. This does not seem to alter the charge transfer significantly. The charge variation upon establishing a current involves mainly the system's conduction band, allowing us to disregard orbitals other than the 2s and three 2p of the oxygen and those in the carbon conduction band, further reducing the computation time. The code was furthermore parallelized.

Since the goal is to study the system in the 'good contact' regime, an infinite tube has been considered. The model is shown in fig.1. Self-consistency is implemented in a long segment of the tube which contains also the ad-atom, while the infinitely extended edges are not modified, and can be projected onto the Hamiltonian by means of a self-energy[15]. The size of the segment in which the potential is varied comprises 22 tube unit cells of a (5,5) tube, which means 440 atoms. The segment is electrostatically shielded from the semi-infinite sides of the tube at the electrodes by two parallel metallic planes at each of its edges, so that charge redistribution does not affect the electrode tube parts. The infinite electrostatic images from the planes have been treated numerically by truncating the series at its 5th element, which gives an almost perfect convergence.

## IV-Results and discussion

Results for the adsorption energy curves are shown in fig.2 for the bridge and top adsorption sites. The adsorption minimum takes place for the bridge position, in agreement with other calculations on graphite[21]. From the figure, the calculated barrier for lateral diffusion is about 0.6 eV, assuming that the diffusion path between bridging sites goes through the on-top site (the energy minimum for the 'center' site is higher than the other two). The calculated perpendicular vibrational frequency at the energy minimum is 0.048 eV, in good agreement with that calculated by ref.[21]. A charge transfer of $0.14e^-$ to the oxygen atom is obtained.

Upon application of a bias between the Fermi levels at the electrodes, a current is established. The adatom acts as a scatterer, resulting in effective charge accumulation at one side of the adatom, and charge depletion on the other side. The magnitude of the total self-consistent charge accumulated per atom is depicted in fig.3, where the color intensity corresponds to the sum of diagonal charges at each atom, being red for electron excess and green for electron deffect. The adsorption position corresponds to a 'diagonal' bridge (*bridge 1* in fig.1), as opposed to the alternative possibility of 'perpendicular' bridge (*bridge 2*). Charge accumulation results in the latter case are qualitatively similar. We see major accumulation at the atoms immediately neighbouring the adatom, and additional very slight accumulation in the long range lengthscale. This last one is a slight net charge accumulation on the emitter side before the impurity, and a slight charge depletion on the other side.

Fig.4 shows the induced charge depletion at the bonds between the adatom and its nearest neighbors when a current is allowed to flow (Vbias=1V). When current flows, antiboding states are populated, while bonding states can partially depopulate, leading to a reduction of the mean charge at the bonds between atoms. This charge redistribution results in a net force on the adatom's core. The direction of the force can be qualitatively deduced from the bond charge depletion graph: if the adsorbate's pseudopotential were constant and localized inside a sphere, the force would be proportional to the integral of the gradient of the charge density inside the sphere. Thus, we see that the force tends to separate the ad-atom from the tube, and also drift it in the direction of the current flow. (In order to see better the direction of the force we have explicitly substracted the ad-atom's intra-atomic charge accumulation density from fig.4, because the integral of its gradient inside the sphere cancels out.)

We thus see that the charge redistribution at the bonds, i.e. the non-diagonal charge, is the strongest effect giving rise to a force on the adsorbate. Indeed, the total contribution of the diagonal charges accumulated along the tube is less than a 10% of the non-diagonal charge contribution.

We have calculated the forces on the adatom for a range of biases, as shown in fig.5. The force is computed easily in the atomic orbital representation, according to eq.(1). The result for the two inequivalent bridging positions can be compared. Since we are neglecting second neighbors interaction, the component of the force out of the plane defined by the adsorbate and its two neighbours is negligible. Also, in adsorption position 2, the parallel component of the force is obviously zero by symmetry. We see that the major component of the force tends to separate the adsorbate from the nanotube, and in the bridge 1 case it slightly pushes it in the direction of electron flow.

The calculated forces are of the order of 0.2 eV/Å for Vbias~2V. This is far from being able to desorb the adsorbate, but can lower the diffusion barrier on one side with respect to the other, enhancing the diffusion probability in one direction. For the case considered, the diffusion barrier is of ~0.6 eV, which would correspond to a thermal diffusion rate of $1s^{-1}$ if the temperature were about 240K. At this temperature, lowering the barrier by 0.05eV implies an increase of one order of magnitude in the diffusion hopping rate. As we have seen, the force component perpendicular to the surface is about 5 times bigger than the parallel component. The perpendicular component affects the diffusion barriers in a symmetric way, and therefore does not induce directional drift. However, it may increase the diffusion rate. If one assumes the diffusion to take place through the saddle points of the potential energy surface, the barriers are effectively lowered by ~0.12 eV at the maximum applied bias of 6V. A molecular dynamics simulation would be useful to clarify the diffusion enhancement. The parallel component affects the diffusion barriers assymetrically, lowering one and raising the other by ~0.05 eV at Vb~6V, which would be able to produce directional diffusion.

However, the mean value of the force is not the only factor affecting diffusion. An important role is played by fluctuations in the mean force, acted upon by inelastic scattering of the tunneling electrons. Such effect is able to produce a cansiderable vibrational heating of the adsorbate by inducing transitions to excited vibrational levels of the adsorbate. The effect of this in the diffusion rate can be very important, as shown in ref.[9]. We can give a fast estimation of the orders of magnitude expected for the inelastic fraction of tunneling electrons scattered by adsorbates at carbon nanotubes. In our case the inelastic matrix elements are mainly due to the variation of the hoppings between the adatom and its neighbors. A typical value of the mean displacement of the vibrational mode is about 1/10 of the bond length. The nearest neighbors hopping is of the order of T~-1eV. Assuming a quadratic dependence of the hopping with distance, this implies that the inelastic matrix elements are of the order of[22] $\delta\varepsilon \sim (dT/dR)\Delta R \sim 0.1eV$. The l.d.o.s. at the Fermi level on our (5,5) tube is about 0.03e/V, and on the impurity it is about 0.1e/V. This implies an inelastic conductance of the order of $\sigma_{inel} \sim 8\frac{e^2}{h}\delta\varepsilon^2 \rho_{subs}\rho_{ads} \sim 10^{-4} G_0$. The fact that there are different paths for the electron to tunnel inelastically means that the inelastic fraction may be of the order of $f \sim 10^{-4}$-$10^{-3}$. Also, the inelastic fraction is expected to vary inversely with the tube's diameter, because it is less likely for the electron to find the impurity when the diameter is bigger. The inelastic fraction value can of course be largely affected by the adsorbate's mass and vibrational frequency. For $f \sim 10^{-3}$. and assuming a typical deexcitation time of $10^{-11}$ s, and a vibrational energy level separation of 0.06 eV, for the sample temperature of 240K used above and 1microAmp current, the vibrational temperature is ~1200K which is a dramatic effect. However, since this result is largely dependent on the parameters employed, we will not discuss further on this point here.

Now we address the question of how the induced force depends on the nanotube's radius. To this end, we have performed the calculation for the adsorbate on bridge position 2, at three different armchair nanotubes of chirality (5,5), (10,10) and (15,15). The force perpendicular to the surface as a function of the current is plotted in fig.6. As a function of bias voltage the current increases faster, the larger the radius of the nanotube, due to a closer opening of subbands around the Fermi level (see inset of fig.6). This leads to a smaller force for the same ammount of current at the (15,15) nanotube with respect to the (5,5) for higher biases. For smaller biases, in the range before the opening of the second subband, the three nanotubes have approximatedly the same current (slightly smaller in the (5,5) and larger in the (15,15), since the adsorbate backscatters electrons more effectively in the former case). The corresponding force at smaller current values is in this case larger for the larger radius tube. The crossover current is roughly inversely proportional to the nanotube radius, thus happening at smaller current values the larger the radius of the nanotube (fig.6).

The reason for a larger force at bigger radii with small applied biases is understood by looking at the non-diagonal local density of states (n.d.l.d.o.s.) in fig.7, as defined in eqs.2-3. A good explanation on the role of the n.d.l.d.o.s. in atomic chemisorption is given by Ishida[23]. In general, when the hopping matrix element between two orbitals is negative (for instance between two s orbitals), positive values of the non-diagonal l.d.o.s. have bonding character, while negative values are antibonding. On the other hand, if the hopping matrix element is positive, as for a p-p$\sigma$ bond, negative n.d. l.d.o.s. implies bonding density, while a positive n.d..l.d.o.s. means antibonding states in the bond at that energy. The induced force on the adsorbate is mainly produced by population of those antibonding states when an electric current is established. From the inset of fig. 7, the largest contribution is from the $p_y$-$\pi$ oxygen-carbon bond, where y is oriented perpendicular to the nanotube's surface. When a bias is applied, an energy stripe around the equilibrium Fermi level becomes repopulated with electrons coming from the emitter side. This results in a higher population of the antibonding states in the $p_y$-$\pi$ bond (positive n.d.l.d.o.s.), which leads to a repulsive force between the adatom and its neighbours. We see from fig.7 that subsequently bigger bias are necessary in order to populate these antibonding levels when going from adsorption on a (15,15) nanotube to adsorption on a (5,5). When going to higher biases the differences in l.d.o.s. near the Fermi level for different radius tubes stop being significant, and approximately the same force is obtained for a given bias in the three cases.

It is important to note that the calculation is performed on a well contacted tube. Such a case has been experimentally achieved[24]. The difficulty to get a good contact is manifested in many experimental works. In case of considerable resistance at the contacts, the currents will be much smaller for the range of biases considered, and the electron distribution will not differ much from the zero current case. Therefore, current induced forces are expected not to play any role in that case. Presumably, only macroscopic electric fields would then be able to exert a force on the ionic adsorbates. Those forces will be strongest in the proximity of a metallic contact. External fields have not been considered in this work, in order to clearly differenciate the effects produced solely by an electric current. The teatment of external electric fields can nonetheless be easily done by simple electrostatic methods[25] up to linear response, or in a more precise way as described in ref.[26]

## V-Conclusions

We have calculated the adsorption energy surface, and the force induced on an O adsorbed on a well contacted (5,5) carbon nanotube by an electric current flowing along the tube, for a range of 6V applied bias. The ab-initio calculated adsorption sites, bond-length, and vibrational frequencies are in agreement with previous calculations. The induced force is of the order of 0.7 eV/Å for the maximum bias applied, with a direction tending to separate the adatom from the substrate, and pulling it in the direction of the current. The main contribution to the force is due to charge redistribution at the adsorbate's bonds, the tunneling electrons populating antibonding levels. There is small charge redistribution also in an extended lenght of the tube, however its effect on the force is minimal. A dependence of the force magnitude with the nanotube radius is found. For biases smaller than the energy of the second subband, larger diameter tubes display a stronger force. The trend is reversed for higher biases and currents, smaller tubes displaying a stronger force for the same ammount of current. These calculated static forces are capable of inducing directionality in the thermal diffusion of the adsorbate; for a 6V bias, the hopping rate in the direction of current flow increases by one order of magnitude with respect to the opposite direction. Inelastic scattering giving raise to force fluctuations also plays a role in enhancing the thermal diffusion, and will be the subject of further work.

## Acknowledgments

NM gratefully acknowledges José Ortega from Autonomous University of Madrid for access to the original fireball code, and Isabel Benito for help in setting up the calculations. Also R.Pérez, F.Flores, M.P.Anantram, Young-Gui Yoon and T.Yamada for useful discussions and comments on the manuscript. We acknowledge Jianping Lu for support in the joint research program between NASA-Ames and the University of North Carolina at Chapel Hill.

**Figures:**

*Fig.1*. Schematic representation of the system. The infinitely long (5,5) nanotube has been considered electrostatically shielded, but for a 55Å region, in the middle of which is located the adsorbate. All electric interactions in the non-shielded region include the electrostatic response of the infinite metallic plates at the sides. The two different adsorption positions considered in the text are shown.

*Fig.2*: Calculated chemisorption curves for atomic oxygen on graphene sheet. Abscissa axis denotes distance from the oxygen nucleus to the center of the line joining the nuclei of its two nearest carbons, in the bridge case, or to the nucleus of its nearest carbon, in the top case. The energies' origin is arbitrary.

*Fig.3*

Fig.3A Induced charge accumulation when a 1V bias is applied (electron flow from right to left takes place). Fig.3B shows the reverse side of the tube for the same case. The scale is not linear, so details of the almost neutral atoms can be shown. The scales shown are from 0.025 e to -0.025 e. The charge on the oxygen is 0.15 e without current and gets to 0.18 e under 1V bias (out of scale).

*Fig.4*. Section plot of nondiagonal charge density accumulated at the bonds after establishing a 1V bias, showing depletion of charge at the bonds. The current flows from right to left, and the adsorption site is bridge 1. The diagonal elements of the total charge density have been substracted in order to better see the influence of the bond's charge on the induced force (see text). The adsorbate and its two neighbors are shown in white, the adsorbate being in the middle of the figure. There is a small area of electron excess over the adsorbate (lighter tones) and an intense electron defect depletion zone at the right hand side bond (darker tones). The level curves are spaced every 0.0015 e/A$^3$. The gray level far from the atoms corresponds to density = 0.

*Fig.5* Mean value of the force induced on the adsorbate as a function of the bias voltage, for two possible bridge adsorption positions. On a bridge perpendicular to the tube axis (bridge 2), the component of the force parallel to the bridge is zero by symmetry. On the 60° rotated bridge (bridge 1), both the parallel component and the perpendicular to bridge and surface component are shown.

Fig.6 Force on the adsorbate on bridge 2, as a function of the current, for three armchair nanotubes of chirality (5,5), (10,10) and (15,15). The inset shows the dependence of current with applied bias for the three cases. $1G_0 = 2e^2/h = (12.9\ \text{Kohm})^{-1}$.

Fig.7 Non-diagonal transmitting local density of states, $\rho^{trans}_{\pi,py}$ between the $p_y$ orbital of the O and one of its neighbouring carbon's $p_y$ orbital, where the y direction is perpendicular to the surface at that point, for the three chiralities considered, and the O adsorbs at bridge 2. The inset shows the non-diagonal densities between the carbon's $p_y$ orbital and the oxygen's s, $p_x$ and $p_y$, for the (5,5) case.

Mingo et al., Fig.1

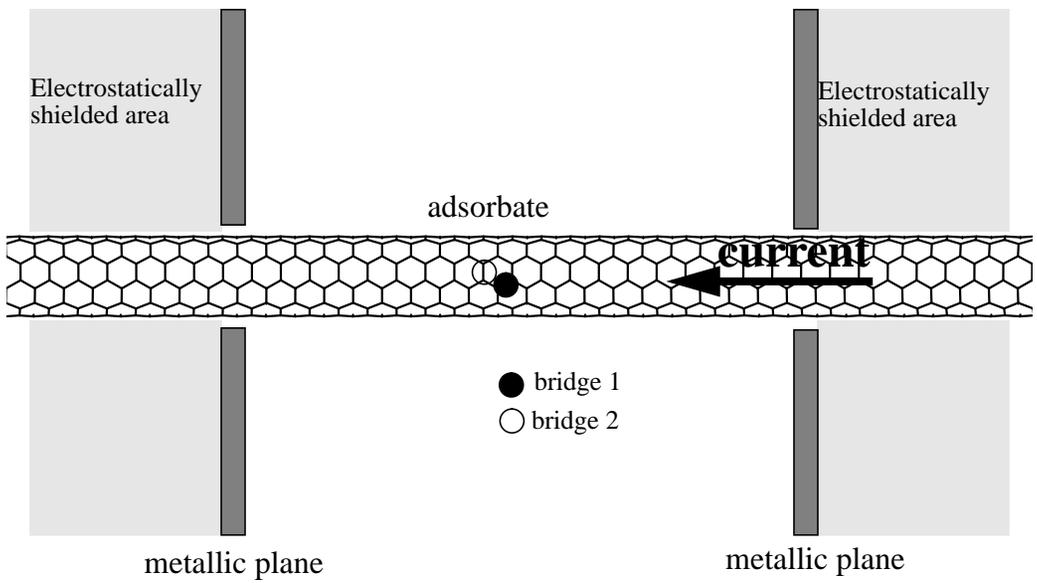

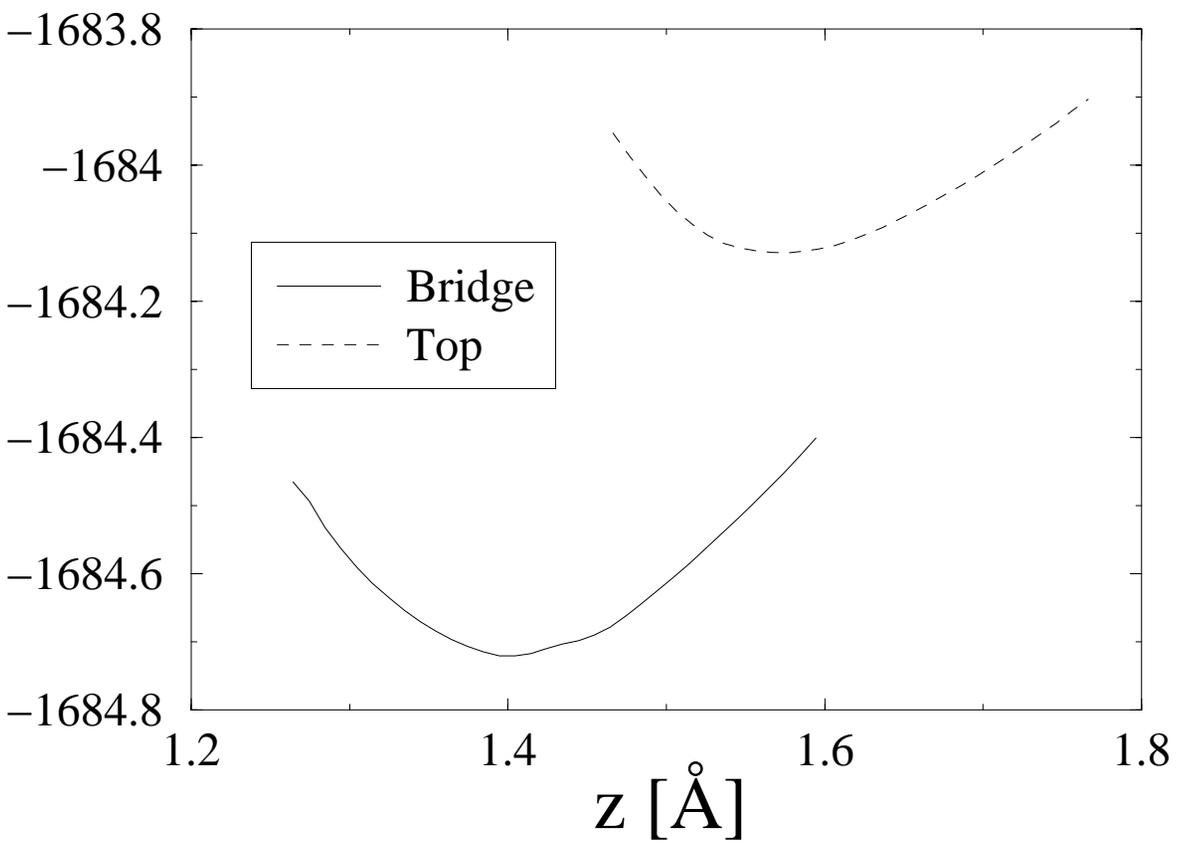

Mingo et al., Fig.2

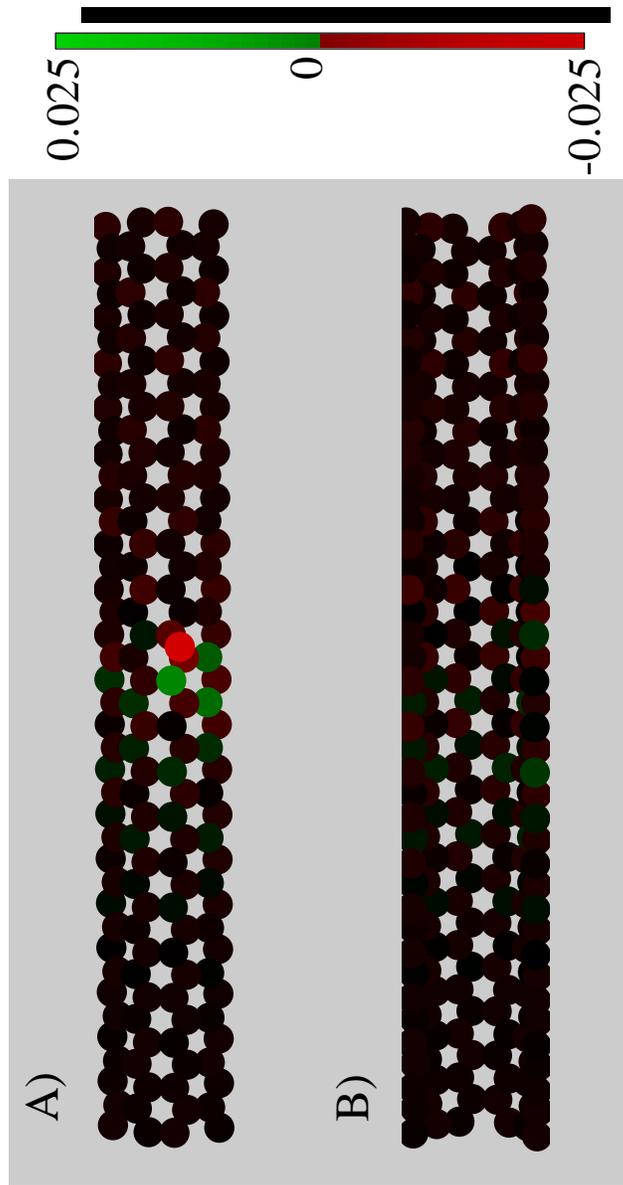

Mingo et al., Fig.3

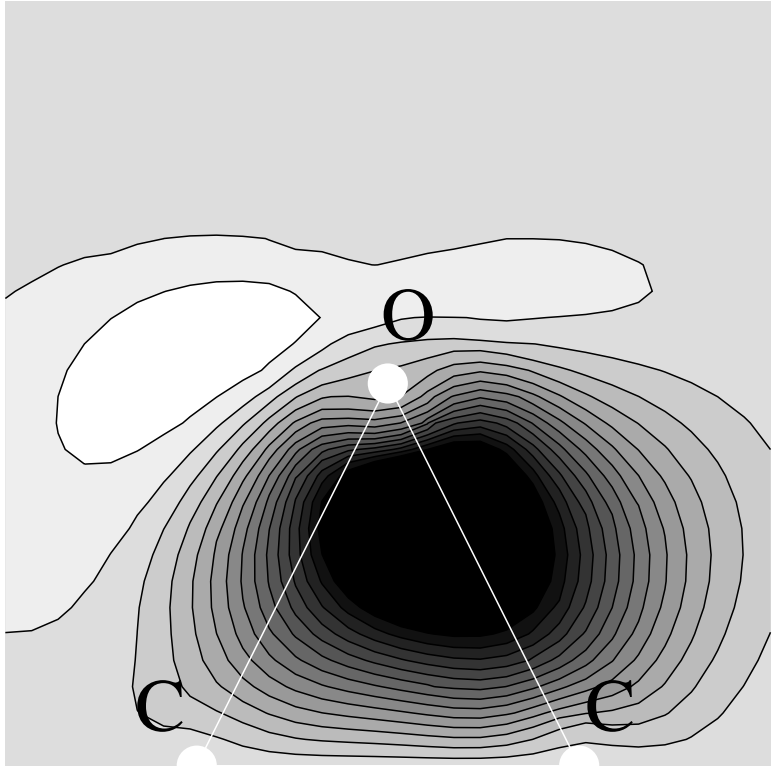

Mingo et al., Fig.4

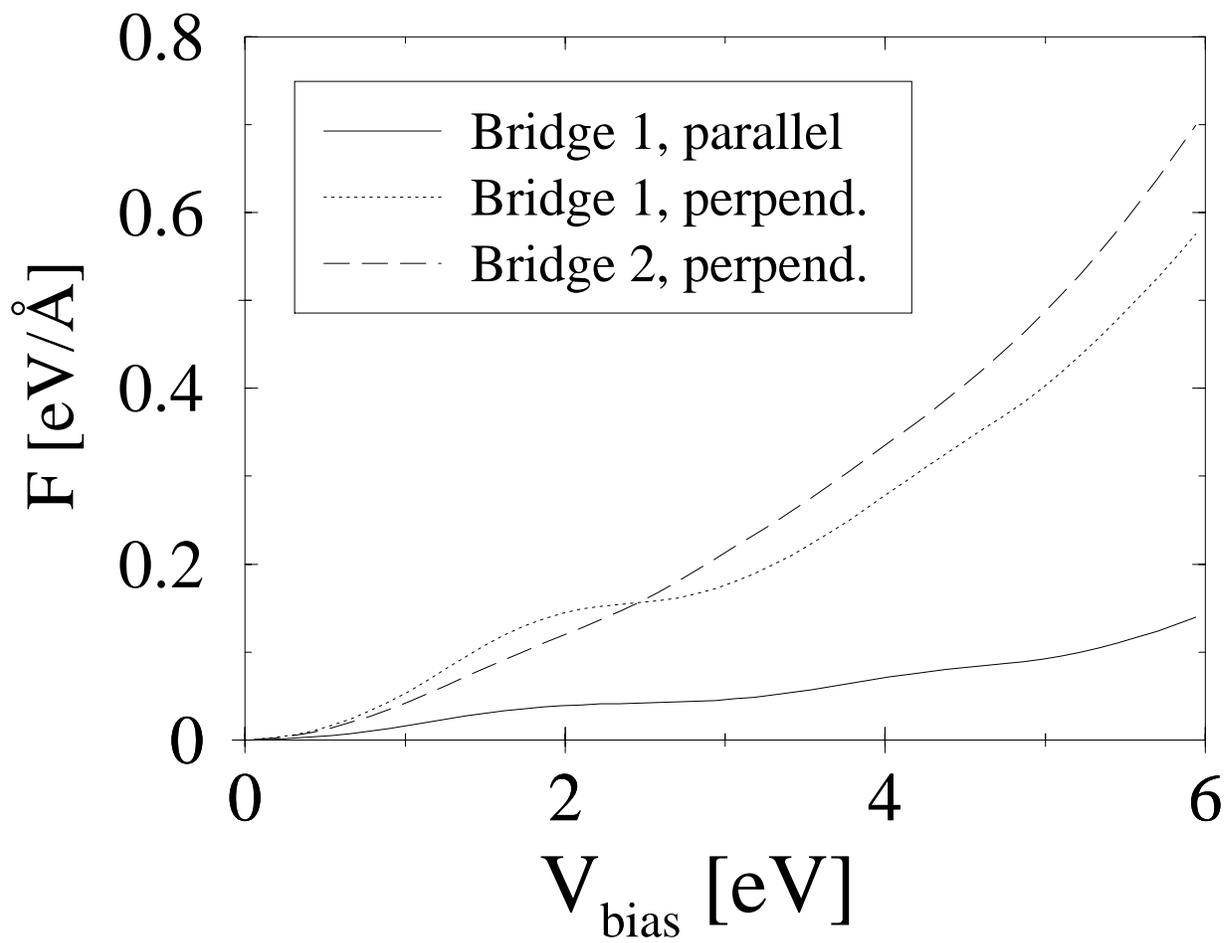

Mingo et al., Fig.5

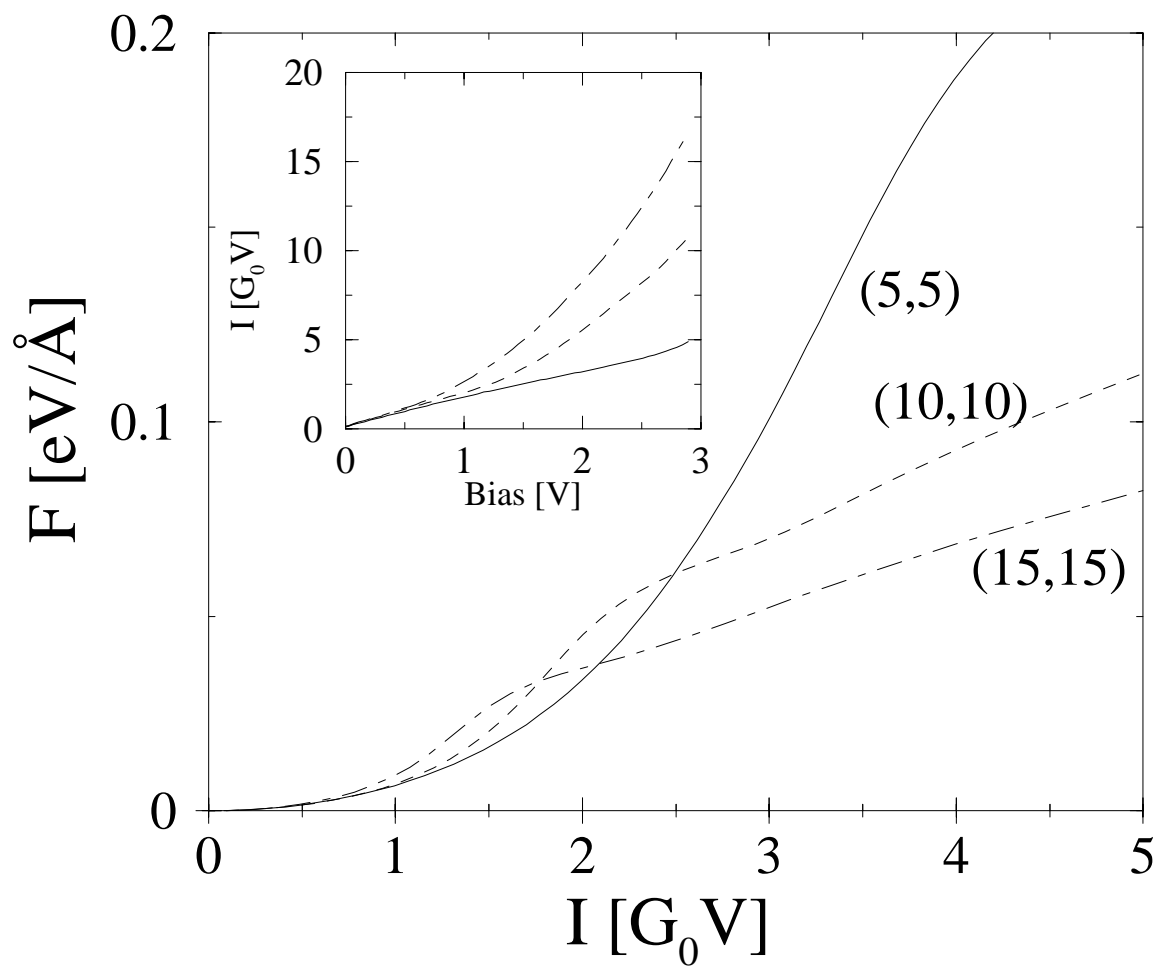

Mingo et al., Fig.6

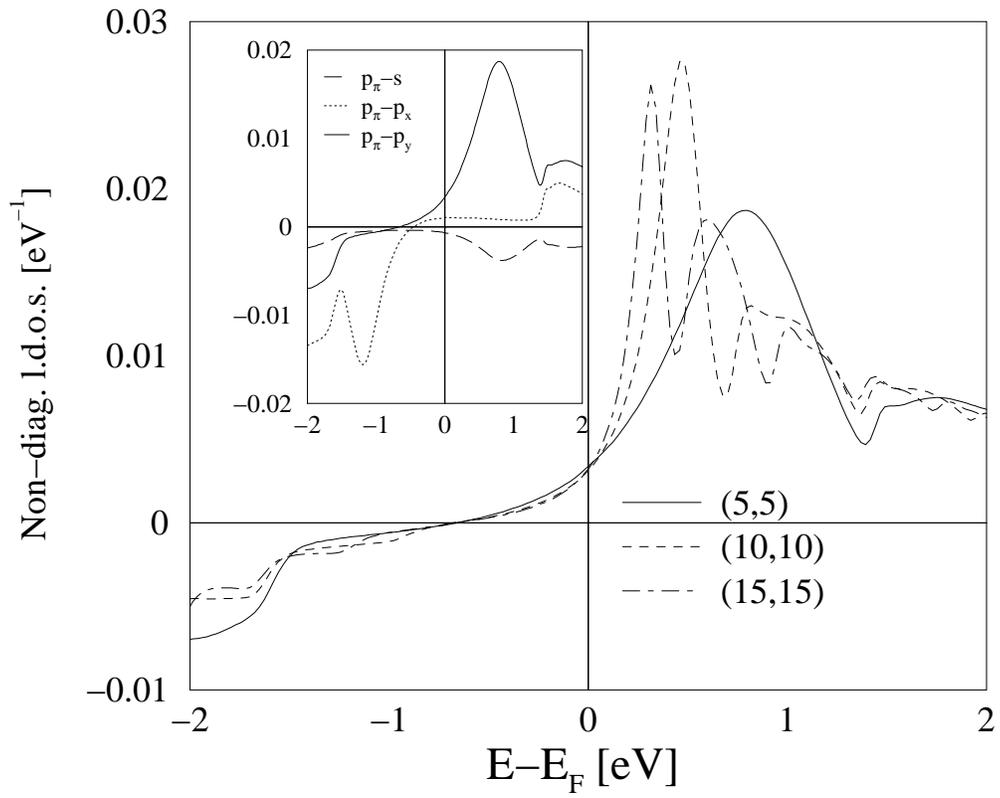

Mingo et al., Fig.7